\documentclass[conference,10pt,letterpaper]{IEEEtran}
\IEEEoverridecommandlockouts
\usepackage{amsopn,amsmath,amssymb,amsfonts,bm,amsthm}
\usepackage{graphicx,subfigure}
\usepackage{enumerate}
\usepackage{cite}
\usepackage{etex}
\usepackage{algorithm,algorithmic}
\usepackage[margin=16mm,top=17mm,bottom=24mm]{geometry}

\def\M{\mathcal M}
\def\P{\mathcal P}

\def\A{\mathcal A}
\def\R{\mathcal R}

\def\C{\mathcal C}
\def\T{\mathcal T}
\def\J{\mathcal J}
\def\K{\mathcal K}

\def\X{\mathcal X}

\newtheorem{Theorem}{\textbf{Theorem}}

\newtheorem{Remark}{\textbf{Remark}}

\newtheorem{Example}{\textbf{Example}}
  {\proof}{\proofend}

\usepackage{color}
\title{Transmission and Scheduling Aspects of Distributed Storage and Their Connections with Index Coding}
\author{Parastoo Sadeghi\\
Research School of Engineering, The Australian National University, Canberra, ACT, 0200, Australia\\\texttt{\{parastoo.sadeghi\}@anu.edu.au}\vspace{-3mm}}

\begin{document}
\maketitle
\begin{abstract}
Index coding is often studied with the assumption that a single source has all the messages requested by the receivers. We refer to this as \emph{centralized} index coding. In contrast, this paper focuses on \emph{distributed} index coding and addresses the following question: How does the availability of messages at distributed sources (storage nodes) affect the solutions and achievable rates of index coding? An extension to the work of Arbabjolfaei et al. in ISIT 2013 is presented when distributed sources communicate via a semi-deterministic multiple access channel (MAC) to simultaneous receivers. A numbers of examples are discussed that show the effect of message distribution and redundancy across the network on achievable rates of index coding and motivate future research on designing practical network storage codes that offer high index coding rates.
\end{abstract}

\section{Introduction}
Index coding \cite{sprintson:matroid:IT:2010} is often studied with the assumption that a single source has all the requested messages. We call this \emph{centralized} index coding.  In practice, messages may be distributed at different sources and the problem of index coding has to be solved with regards to the availability of messages at different sources. This leads to \emph{distributed} index coding. In this context, this paper addressed the following question
\begin{itemize}
\item How does the availability of messages at distributed sources affect the solutions and achievable rates of index coding?
\end{itemize}


In addition, index coding is often studied with the assumption of a perfect (noise-free) unit-capacity link between a single source and receivers. In the case of multiple distributed sources, one has to take into account the transmission medium when solving the index coding problem. In this paper, we consider a semi-deterministic (noise-free) multiple access channel (MAC) between distributed sources and a set of \emph{simultaneous} receivers and study the following questions 
\begin{itemize}
\item How does the transmission and scheduling in the MAC affect the capacity region of index coding?
\item How distributed storage can improve the capacity region of index coding?
\end{itemize}

Finally, we touch upon the impact of coded storage on the achievable rates in distributed index coding.

To the best of author's knowledge, the only other work that studies index coding with multiple sources is \cite{ong:multisource:arxiv}. However, there are four fundamental differences between our work and \cite{ong:multisource:arxiv}. Firstly, the authors in \cite{ong:multisource:arxiv} assume that receivers have \emph{unipriors}, which means that they only know one unique message a priori. We do not require this in our model. Secondly, they assume that each receiver may require more than one message. Here we assume that receiver $j$ requires message $M_j$, which can be extended to more general cases. Thirdly, they use a graph theoretical approach to provide bounds on the multiple source \emph{linear} index coding rates. Here, we do not limit ourselves to linear coding and also adopt the method of composite message encoding in \cite{ong:multisource:arxiv} to directly derive achievable rate regions. Finally, the authors in \cite{ong:multisource:arxiv} assume that there are noiseless orthogonal \emph{bit pipes} between each source to the receivers, which will somewhat simplify the problem. Here, we do not assume such channel model and work with a MAC where transmissions from sources can interfere with each other.

The distributed index coding problem that we consider in this paper can have applications in wireless storage networks where \emph{hot content} (such as popular video) need to be provided at high rates to a set of wireless clients who already have a priori knowledge of some video files in their caches. However, we note that the system setup in this paper is fundamentally different from recent work on coded caching \cite{caching:fundamental:it:2014} in a number of ways. First, we do not optimize for the \emph{placement} of messages at the end receivers under worst case unique message requests. In addition, we do not assume a single server containing all messages is involved in the \emph{delivery phase} transmissions. Thirdly, we do not consider receiver cache-size versus server delivery rate tradeoffs.

\section{Problem Formulation and Background}

Consider $K$ distributed sources who wish to communicate $N$ messages, where message $M_j \in [1:2^{nR_j}]$, $j \in [1:N]$, is intended for respective receiver $j$. Each source $k\in[1:K]$, has access to subset of messages $M_{\T_k}$, where $\T_k \subseteq [1:N]$ with the condition that $\bigcup_{k=1}^K \T_k = [1:N]$. Each receiver $j\in [1:N]$ has prior knowledge of a subset of messages $M_{\A_j}$, where $\A_j \subseteq [1:N]\setminus \{j\}$. Based on available messages $M_{\T_k}$, each source $k$ can send a sequence of $n$ symbols $x_k^n$ over a semi-deterministic noise-less discrete memoryless MAC where the channel output at time index $i$ solely depends on channel  inputs at time $i$ through a known and fixed function $y_i = f(x_{1,i}, x_{2,i}, \cdots, x_{K,i})$, which is simultaneously received by all receivers. Based on the received sequence $y^n$, each receiver $j$ finds an estimate $\hat M_j$ of the message $M_j$. Note that this distributed MAC index coding problem is fully characterized by source message index sets $\T_k$, $k \in [1:K]$, receiver side information index sets $\A_j$, $j \in [1:N]$ and channel function $y_i = f(x_{1,i}, x_{2,i}, \cdots, x_{K,i})$. 

We define a $(2^{nR_1}, 2^{nR_2}, \cdots, 2^{nR_N},n)$ code for index coding by $K$ encoders of form $x_k^n(m_{\T_k})$ and $N$ decoders $\hat m_j(y^n, m_{\A_j})$. We assume that the message tuple $(M_1,M_2, \cdots, M_N)$ is uniformly distributed over $[1:2^{nR_1}]\times [1:2^{nR_2}]\times\cdots\times [1:2^{nR_N}]$. The average probability of error is defined as $P_e^{(n)}=\Pr((\hat M_1, \cdots, \hat M_N) \neq (M_1, \cdots, M_N))$. A rate tuple $(R_1, R_2, \cdots, R_N)$ is achievable if there exists a sequence of $(2^{nR_1}, 2^{nR_2}, \cdots, 2^{nR_N},n)$ codes such that $\lim_{n \to \infty} P_e^{(n)} = 0$. The capacity region $\C$ of the index coding problem is the closure of all achievable rate tuples $(R_1, R_2, \cdots, R_N)$.

The goal is to find an achievable rate region of distributed index coding and the storage and transmission schemes that can achieve it. Before, addressing this problem, we review some known results in the literature.
\subsection{Centralized Index Coding}

In the centralized index coding, there is a single server ($K=1$) which has all messages. Hence, $\T_1 = [1:N]$. Also, a noiseless unit capacity common link is assumed between the source and the receivers. 

The following example of centralized index coding was presented in \cite{capacity:region}, which will form the basis for new examples of distributed index coding discussed later in the paper. 

\begin{Example}\label{exmp:central:1}
A single source has 4 messages, $\{1,2,3,4\}$, and receiver $j \in [1:N]$ wants message indexed by $j$ and has messages indexed by $\A_1 = \{4\}$, $\A_2 = \{3,4\}$, $\A_3 = \{1,2\}$, $\A_4 = \{2,3\}$, respectively, which is represented compactly as the following collection of $\{(j|\A_j)\}_{j=1}^{N}$
$$\{(1|4), (2|3,4), (3|1,2), (4|2,3)\}.$$
\end{Example}

An inner bound for the centralized index coding is shown to be achieved through \emph{composite-coding}. In composite-coding the single sender is replaced by $2^N-1$ ``virtual'' encoders, each encoding a non-empty subset $M_\J$, $\J\subseteq [1:N]$, of the message tuple $(M_1, \cdots, M_N)$ into a composite message $W_\J$  at rate $S_\J$ in a ``flat'' manner (as opposed to layered superposition coding). This part is referred to as dual index coding problem and its capacity region is the set of rate tuples $(R_1, R_2, \cdots, R_N)$ that satisfy
\begin{align}\label{eq:dual:region:1}
\sum_{j \in \J} R_j \leq \sum_{\J'\subseteq[1:N]:\J' \cap \J \neq \emptyset}S_{\J'}
\end{align}
for all $\J \subseteq [1:N]$. 

Having received all composite messages $W_\J$, $\J\subseteq [1:N]$, receiver $j$ can employ simultaneous non-unique decoding together with its side information to decode any subset $\K_j$ of messages where the capacity region $\R(\K_j|\A_j)$ is defined by
\begin{align}\label{eq:dual:region:2}
\sum_{j \in \J} R_j <\sum_{\J'\subseteq\K_j\cup\A_j:\J' \cap \J \neq \emptyset}S_{\J'}
\end{align}
for all $\J \subseteq \K_j\setminus \A_j$. By considering all possible message subsets $\K_j$, $\K_j\subseteq [1:N]:j\in \K_j$ that include the desired message $j$, and all composite message indices that are relevant to the message subset $\K_j$, $\J'\subseteq\K_j\cup\A_j:\J' \cap \J \neq \emptyset$, achievability of the following rate region is argued in \cite{capacity:region}.

\begin{Theorem}[Centralized Composite-Coding Inner Bound\cite{capacity:region}] \label{th:centralized} A rate tuple $(R_1, R_2, \cdots, R_N)$ is achievable for centralized index coding problem $\{(j|\A_j)\}$ if
\begin{align}\label{eq:index:region:1}
(R_1, R_2, \cdots, R_N) \in \bigcap_{j\in[1:N]}\quad\bigcup_{\K_j\subseteq [1:N]:j\in \K_j} \R(\K_j|\A_j)
\end{align}
for some $(S_\J: \J \subseteq [1:N])$ such that $\sum_{\J:\J \subsetneq \A_j} S_\J \le 1$ for all $j \in [1:N]$.
\end{Theorem}

\begin{figure*}[t]
\centering
\includegraphics[width=0.8\linewidth]{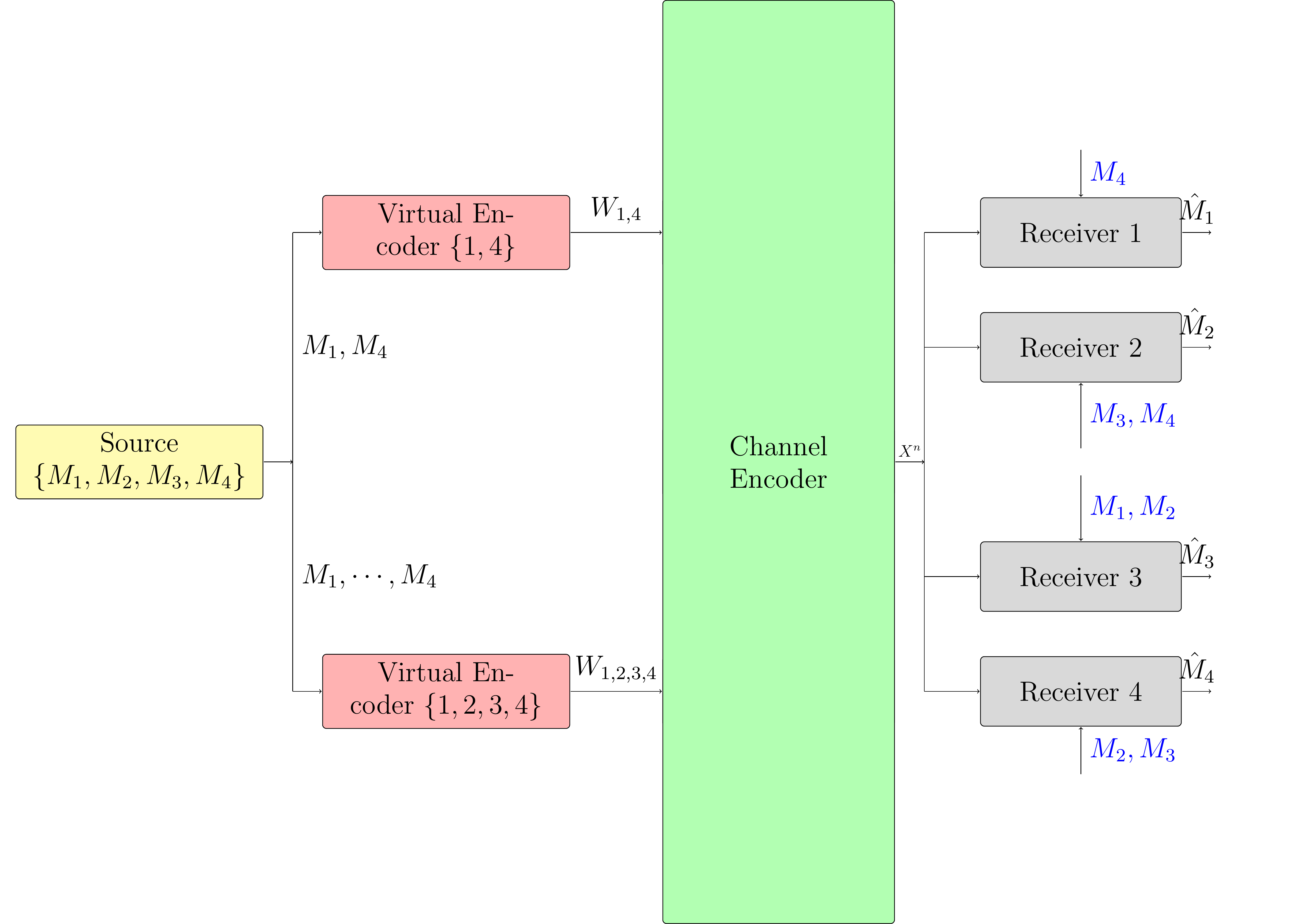}
\caption{The schematic for virtual encoding and channel encoding of composite messages for Example \ref{exmp:central:1:RR}.}
\label{fig:cent:1234}
\end{figure*}
\begin{figure*}[t]
\centering
\includegraphics[width=0.8\linewidth]{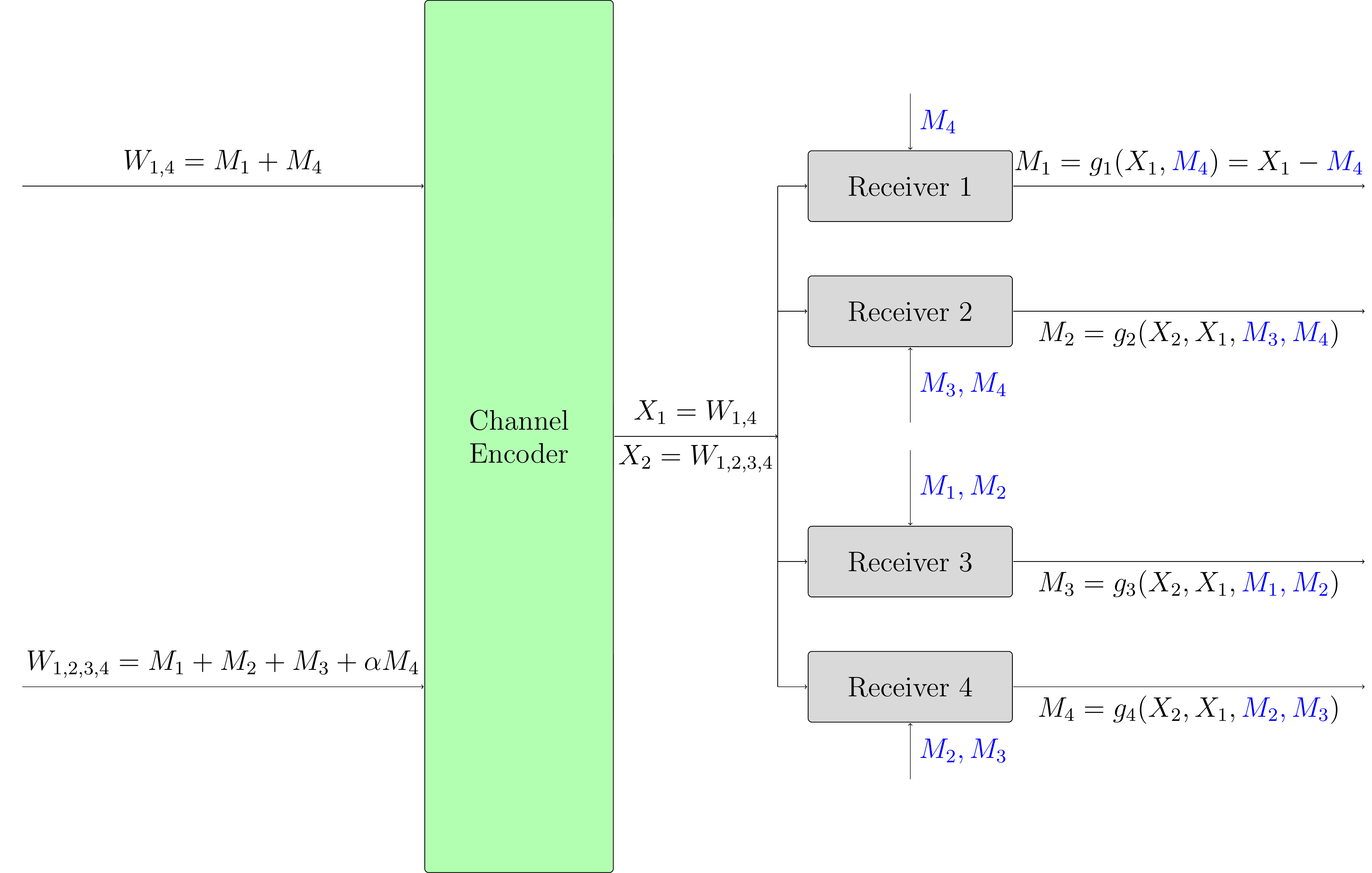}
\caption{The overall schematic of decoding messages for Example \ref{exmp:central:1:RR}. Channel input alternates between $X_1$ and $X_2$ in two successive channel uses and $g_j$ is the decoding function of user $j$.}
\label{fig:cent:1234:decod}
\end{figure*}

\begin{Example}\label{exmp:central:1:RR}
The capacity inner bound of Example \ref{exmp:central:1} is determined by setting $\K_1 = \{1\}, \K_2 = \{1,2\}, \K_3 = \{3,4\},\K_4 = \{1,4\}$ and the set of constraints
\begin{align}\label{eq:central:RR:example}
R_1&\le S_{1,4},\\R_2 &\le S_{1,2,3,4}\\
R_1+R_2 &\le S_{1,4}+S_{1,2,3,4}, \\
R_3&\le S_{1,2,3,4}\\ R_3+R_4 &\le S_{1,4}+S_{1,2,3,4}\\
R_1+R_4 &\le S_{1,4}+S_{1,2,3,4}, \\
S_{1,4}+S_{1,2,3,4}&\le 1
\end{align}
where after Fourier-Motzkin elimination results in
\begin{align}\label{eq:central:RR:example:1}
R_1+R_2 &\le 1,\quad R_1+R_3 \le 1\\
R_1+R_4 &\le 1, \quad R_3+R_4 \le 1
\end{align}
which was shown in \cite{capacity:region} to be indeed tight. Note that for notational convenience in examples in this paper, we replace notations of the form $S_{\{1,4\}}$ with $S_{1,4}$. The encoding schematic is shown in Fig. \ref{fig:cent:1234}. The rates $R_j = 0.5$ and hence sum rate of $\sum_j R_j = 2$ can be achieved by alternately encoding $X_1 = W_{1,4} = M_1+M_4$ and $X_2 = W_{1,2,3,4} = M_1+M_2+M_3+\alpha M_4$, where the messages and coefficient $\alpha \neq 1$ are chosen from a finite field $\mathbb F_q$.  The decoding is shown in Fig. \ref{fig:cent:1234:decod}.
\end{Example}

\section{An Achievable Rate Region for Distributed Index Coding}
In this section, we propose an extension of the centralized composite-coding inner bound for distributed index coding. We highlight the main differences as we build the ingredients of the inner bound.

The main difference is that each source $k$ has now access to only $2^{T_k}-1$ composite messages where $T_k$ is the size of $\T_k \subseteq [1:N]$. The composite messages at source $k$ are of the form $W_\J$ corresponding to  a given subset $M_\J$ of the message tuple $(M_1, \cdots, M_N)$ for $\J\subseteq \T_k$. Denote by $\P'(\X) = \P(\X)\setminus \emptyset$, where $\P$ is the power set of $\X$. Then the set of all possibly computable composite messages in the network is $$ \P' = \{W_\J: \quad \J \in\bigcup_{k=1}^K \P'(\T_k)\}.$$

Each distributed source $k$ sends a sequence $x_k^n$, which encodes composite messages $W_\J$ for all $\J\subseteq \T_k$. Then $y^n$ will be simultaneously received by all receivers according to $y_i = f(x_{1,i}, x_{2,i}, \cdots, x_{K,i})$. Let $S_\J$ denote the rate with which composite message $W_\J$,  $\J \in \P'$ is transmitted across the MAC. A rate tuple $(S_\J:\J \in \P')$ is achievable if it belongs to the capacity region of this MAC denoted by $\M$. Note that when the messages of sources are overlapping, then there will be some virtual distributed encoders with identical composite messages.

\begin{figure*}[t]
\centering
\includegraphics[width=0.8\linewidth]{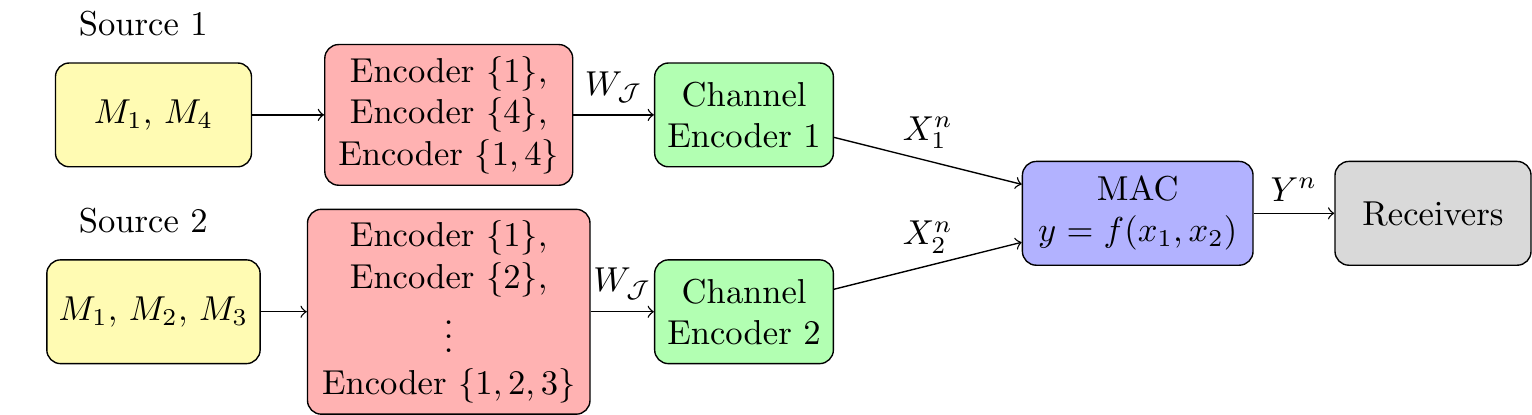}
\caption{The schematic for virtual encoding and channel encoding of composite messages for Examples \ref{exmp:dist:14:1234:intro} and \ref{exmp:dist:14:123}. Note that $W_{\{1\}}$ is common between two corresponding virtual encoders at sources 1 and 2. }
\label{fig:dist:14:123}
\end{figure*}

\begin{Example}\label{exmp:dist:14:1234:intro}
An example of virtual encoders with $K=2$ sources, $\T_1 = \{1,4\}$, and $\T_2 = \{1, 2, 3\}$ is shown in Fig. \ref{fig:dist:14:123} with $$\P' = \{\{1\}, \{4\}, \{1,4\}, \{2\}, \cdots, \{1,2,3\}\}.$$  Note that $W_{\{1\}}$ is common between two corresponding virtual encoders at sources 1 and 2. 
\end{Example}

The rest of the coding scheme follows that of \cite{capacity:region} with appropriate modifications as we now show. The capacity region of distributed dual index coding problem is the set of rate tuples $(R_1, R_2, \cdots, R_N)$ that satisfy
\begin{align}\label{eq:dist:dual:region:1}
\sum_{j \in \J} R_j \leq \sum_{\J'\in \P':\J' \cap \J \neq \emptyset}S_\J
\end{align}
for all $\J \subseteq [1:N]$. Compared to \eqref{eq:dual:region:1}, we take into account only rates  $S_\J$, $\J \in \P'$ of composite messages that are possible to compute in the network. Similarly, $\R(\K_j|\A_j)$ in \eqref{eq:dual:region:2} should be modified as

\begin{align}\label{eq:dist:dual:region:2}
\sum_{j \in \J} R_j <\sum_{\J'\in (\P(\K_j\cup\A_j)\cap \P'):\J' \cap \J \neq \emptyset}S_\J
\end{align}
for all $\J \subseteq \K_j\setminus A_j$. By considering all possible message subsets $\K_j$, $\K_j\subseteq [1:N]:j\in \K_j$ that include the desired message $j$, and all possible composite message indices that are both relevant to the message subset $\K_j$ and can be computed in the network, $\J'\in (\P(\K_j\cup\A_j)\cap \P'):\J' \cap \J \neq \emptyset$, the following rate region is achievable for distributed index coding.
\begin{Theorem}[Distributed  Composite-Coding Inner Bound] A rate tuple $(R_1, R_2, \cdots, R_N)$ is achievable for distributed index coding problem with $\{\T_k\}_{k=1}^K$, $\{(j|\A_j)\}_{j=1}^N$, $y_i = f(x_{1,i}, x_{2,i}, \cdots, x_{K,i})$ and MAC capacity region $\M$ if 
\begin{align}\label{eq:index:region:1}
(R_1, R_2, \cdots, R_N) \in \bigcap_{j\in[1:N]}\quad\bigcup_{\K_j\subseteq [1:N]:j\in \K_j} \R(\K_j|\A_j)
\end{align}
for some $\J^* \subseteq \P'$ such that for all $j \in [1:N]$ and for all $\tilde{\J} \subseteq \J^* : \exists \J \in \tilde{\J}: \K_j \cap \J \neq \emptyset$ we have $\sum_{\J\in \tilde{\J}: \J \subsetneq \A_j} S_\J$ belong to the MAC capacity region $\M$.


\end{Theorem}\label{main:theorem}


\section{Examples and Insights}
\begin{figure*}[t]
\centering
\includegraphics[width=1\linewidth]{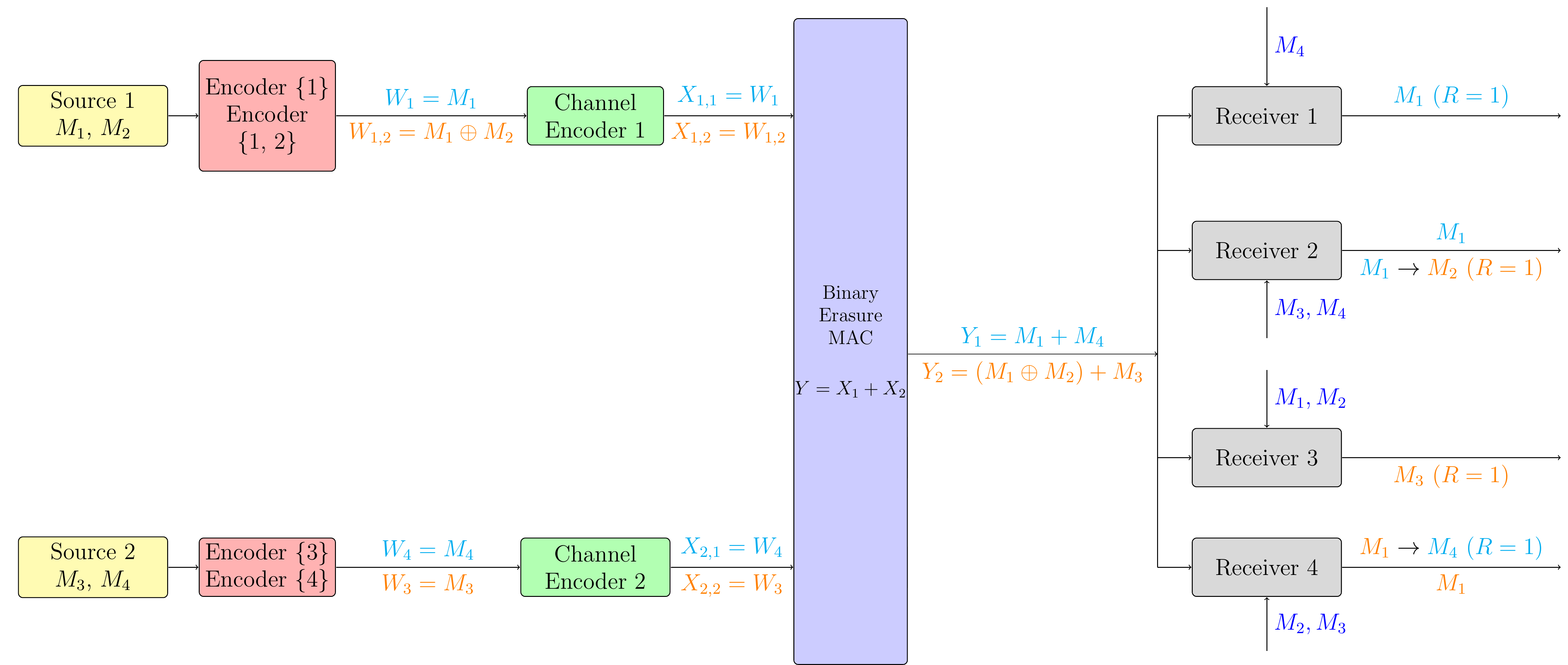}
\caption{The overall schematic of decoding messages for Example \ref{exmp:dist:12:34}. The output of virtual encoders, channel encoders and the channel at two successive channel uses are shown vertically. Note that the shown rates next to each desired message need to be averaged across two channel uses. Hence, $R_1 = R_2=R_3=R_4 = 0.5$.}
\label{fig:dist:12:34:decod}
\end{figure*}

To illustrate the above rate region and better understand the impact of availability of messages at distributed sources, we consider a set of examples of somewhat progressive complexity. Unless otherwise stated, we consider a noiseless binary erasure MAC of the form $y=\sum_{k=1}^Kx_k$, where $x_k \in \X =\{0,1\}$ and summation is in real domain, the capacity region of which is well known \cite{Cover2006}.
\begin{Example}\label{exmp:dist:12:34}
Building on Example \ref{exmp:central:1}, assume the same receivers' message requests and has sets and consider $K=2$ sources with $\T_1 = \{1,2\}$, and $\T_2 = \{3,4\}$ and $y=x_1+x_2$. Note that there is no commonality between source messages and $$\P' = \{\{1\}, \{2\}, \{1,2\}, \{3\}, \{4\}, \{3,4\}\}.$$ We set $\K_1 = \{1\}, \K_2 = \{1,2\}, \K_3 = \{3\},\K_4 = \{1,4\}$  and obtain the following constraints on the index coding rates
\begin{align}\label{eq:dist:RR:example1}
R_1&\le S_{1},\\
R_2 &\le S_{1,2}\\
R_1+R_2 &\le S_1+S_{1,2}\\
R_3 &\le S_3\\
R_4&\le S_4\\
R_1+R_4 &\le S_1+S_{1,2}+S_4
\end{align}
Therefore, in Theorem \ref{main:theorem}, we have $\J^* = \{\{1\}, \{1,2\}, \{3\}, \{4\}\}$. To obtain the constraints on rates $S_j$, we proceed as follows. For $j=1$, excluding element $\{4\}$ from $\J^*$, $\tilde{\J} =  \{\{1\}\}$, $\tilde{\J} =  \{\{1\}, \{1,2\}\}$, $\tilde{\J} =  \{\{1\}, \{3\}\}$, and $\tilde{\J} =  \{\{1\}, \{1,2\}, \{3\}\}$ are relevant. For $j=2$, excluding elements $\{3\}$ and $\{4\}$  from $\J^*$, $\tilde{\J} =  \{\{1\}\}$, $\tilde{\J} =  \{\{1,2\}\}$ and $\tilde{\J} =  \{\{1\}, \{1,2\}\}$ are relevant. Similarly, for $j=3$, $\tilde{\J} =  \{\{3\}\}$ and $\tilde{\J} =  \{\{3\}, \{4\}\}$ are relevant. Finally, for $j=4$, $\tilde{\J} =  \{\{4\}\}$, $\tilde{\J} =  \{\{1\}, \{4\}\}$ and $\tilde{\J} =  \{\{1\}, \{1,2\},\{4\}\}$ are relevant. Hence, we obtain the following effective constraints:
\begin{align}\label{eq:dist:RR:example11}
S_{1}+S_{1,2} &\le 1\\
S_1+S_{1,2}+S_3 &\le 1.5\\
S_{3}+S_{4} &\le 1\\
S_{1}+S_{1,2}+S_4 &\le 1.5
\end{align}
where after Fourier-Motzkin elimination results in
\begin{align}\label{eq:dist:RR:example1:1}
R_1+R_2 &\le 1,\quad \quad R_1+R_2+R_3 \le 1.5,\\
R_3+R_4 &\le 1,\quad \quad R_1+R_2+R_4 \le 1.5.
\end{align}


To verify achievability, consider Fig. \ref{fig:dist:12:34:decod}. Set $W_1 = M_1$ at rate $S_1=1$,  $W_{1,2} = M_1\oplus M_2$ at rate $S_{1,2}=1$, $W_3 = M_3$ at rate $S_3=1$ and $W_4 = M_4$ at rate $S_4=1$. In channel use 1, receivers 1 and 2 decode $M_1$ at rate $R=1$ from $Y_1 = M_1+M_4$ since $M_4$ is known to them. In channel use 2, receivers 3 and 4 decode $M_3$ and $M_1$, respectively at rates $R=1$ from $Y_2 = (M_1\oplus M_2)+M_3$ using their known messages. Then, receiver 4, uses $M_1$ back in $Y_1$ to decode its desired message $M_4$ at rate $R=1$.  Therefore, average rates of $R_1 =R_2=R_3 = R_4 = 0.5$ are achievable over two channel uses. 
\end{Example}

\begin{figure*}[t]
\centering
\includegraphics[width=\linewidth]{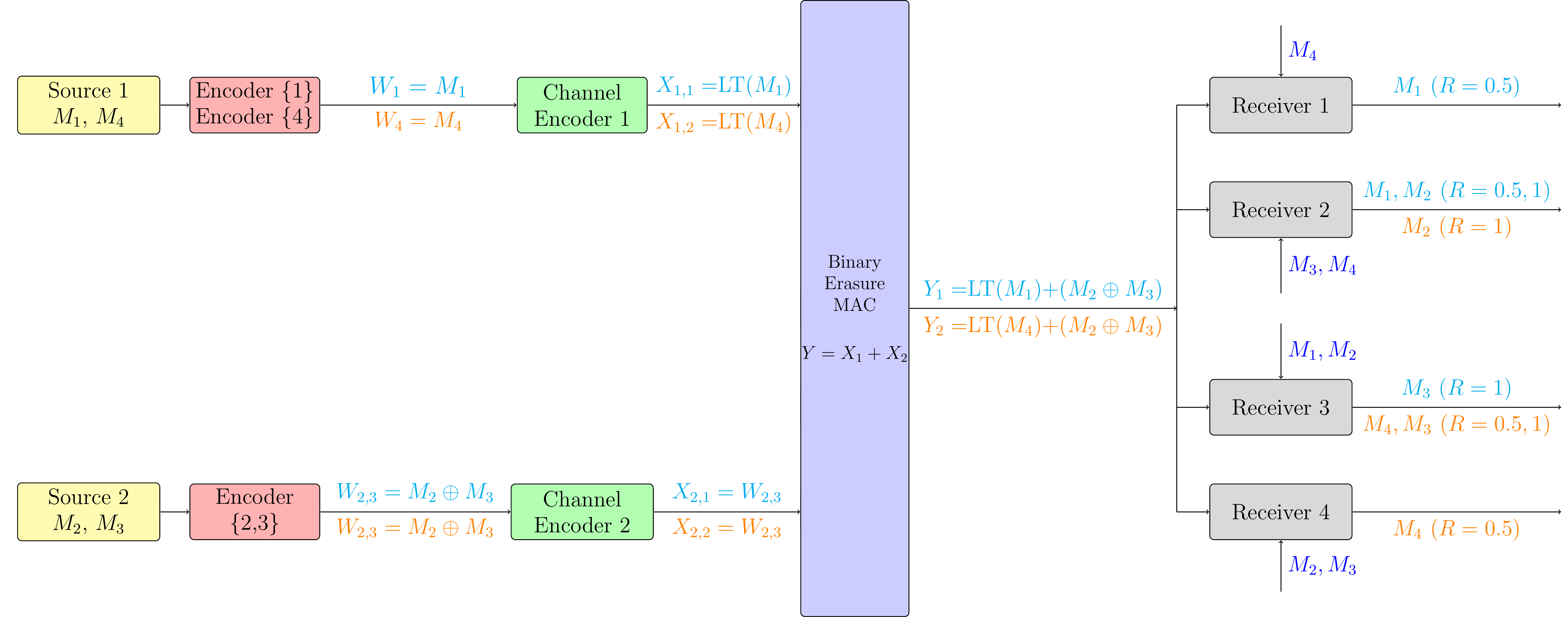}
\caption{The overall schematic of decoding messages for Example \ref{exmp:dist:14:23}.}
\label{fig:dist:14:23:decod}
\end{figure*}

\begin{Example}\label{exmp:dist:14:23}
Building on Example \ref{exmp:dist:12:34}, consider $K=2$ sources with $\T_1 = \{1,4\}$, and $\T_2 = \{2,3\}$. Note that there is no commonality between source messages and $$\P' = \{\{1\}, \{4\}, \{1,4\}, \{2\}, \{3\}, \{2,3\}\}.$$ We set $\K_1 = \{1\}, \K_2 = \{1,2\}, \K_3 = \{3,4\},\K_4 = \{1,4\}$  and obtain the following constraints on the index coding rates
\begin{align}\label{eq:dist:RR:example2}
R_1&\le S_{1}+S_{1,4},\\R_2 &\le S_2+S_{2,3}\\
R_1+R_2&\le S_{1}+S_{1,4}+S_2+S_{2,3}\\
R_3&\le S_3+S_{2,3},\\
R_3+R_4&\le S_{3}+S_{2,3}+S_{1,4}\\
R_4 &\le S_{1,4}\\
R_1+R_4&\le S_{1}+S_{1,4}
\end{align}
To obtain the constraints on rates $S_\J$, we proceed as follows.  For $j=1$, the effective MAC constraints are
$S_1+S_{1,4} \le 1$ and $S_1+S_{1,4}+S_2+S_3+S_{2,3} \le 1.5$.  For $j=2$, the effective MAC constraints are
$S_2+S_{2,3} \le 1$ and $S_1+S_{1,4}+S_2+S_{2,3} \le 1.5$ and so on. Overall, the effective constraints are
\begin{align}\label{eq:dist:RR:example11}
S_1+S_{1,4}+S_2+S_3+S_{2,3} &\le 1.5\\
S_{2}+S_{2,3} &\le 1\\
S_3+S_{2,3} &\le 1\\
S_{3}+S_{2,3}+S_{1,4}&\le 1.5\\
S_{1}+S_{1,4} &\le 1
\end{align}
where after Fourier-Motzkin elimination results in
\begin{align}\label{eq:dist:RR:example2:1}
R_1+R_2+R_3 &\le 1.5,\\
R_1+R_2+R_4 &\le 1.5\\
R_1+R_4 &\le 1\\
R_2 &\le 1\\
R_3 &\le 1
\end{align}
As shown in Fig. \ref{fig:dist:14:23:decod}, average rates $R_1 = R_4 = 0.25$ and $R_2=R_3 = 1$ are achievable by setting $S_1 = S_4 = 0.5$ and $S_{2,3} = 1$, resulting in the increased sum rate of $R_1+R_2+R_3+R_4  = 2.5$, compared to Examples \ref{exmp:central:1:RR} and \ref{exmp:dist:12:34}.\footnote{Throughout the paper, use of proper erasure block channel codes is assumed. In the figures, notations such as LT($M_j$) refer to erasure block coding of message $M_j$.} Note that the relaxed constraints on $R_2$ and $R_3$ come from the fact that receivers 2 and 3 know each other messages and hence, their presence in one source node facilitates transmission of a suitable composite message at rate $S_{2,3} = 1$ for both receivers.
\end{Example}

The next aspect that we wish to explore is the effect of repetition on the rate regions of distributed index coding.

\begin{figure*}[t]
\centering
\includegraphics[width=\linewidth]{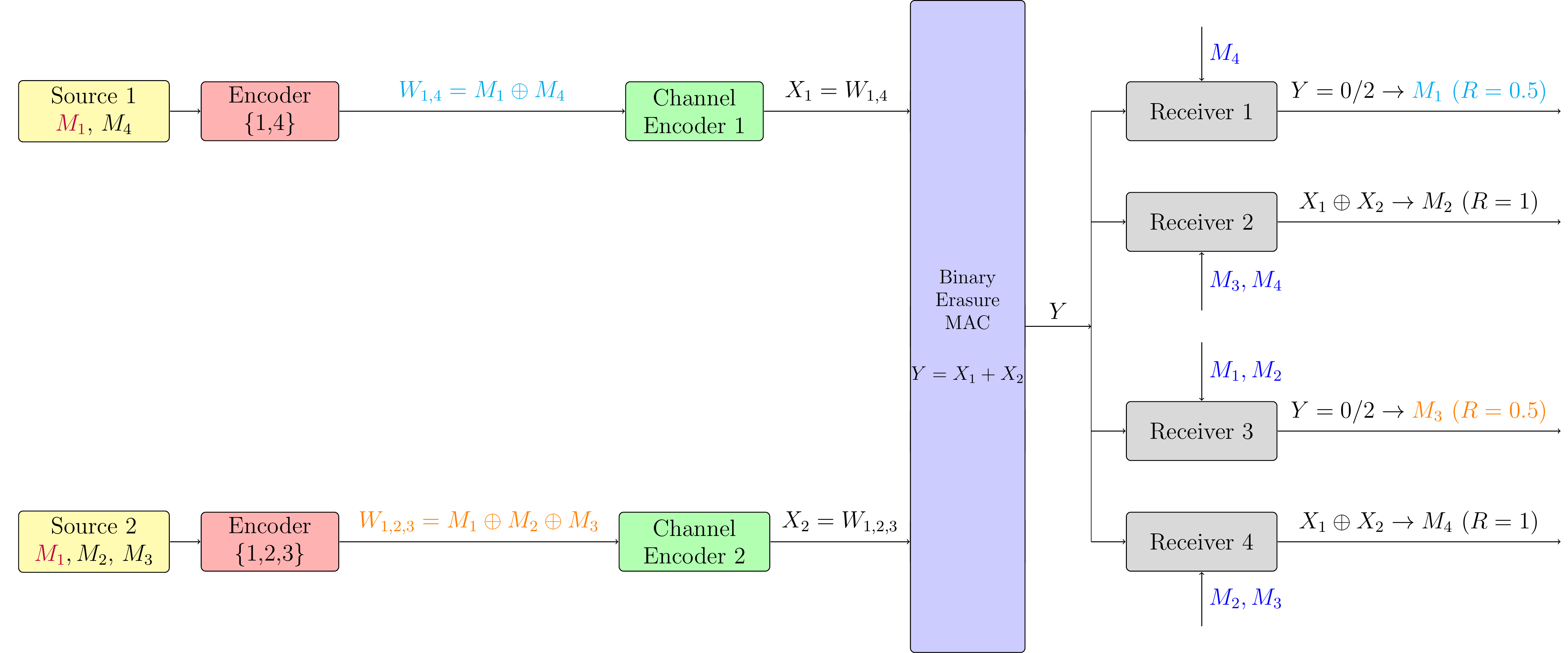}
\caption{The overall schematic of decoding messages for Example \ref{exmp:dist:14:123}. Note that receivers 1 and 3 can only decode when MAC output $Y$ is 0 or 2.}
\label{fig:dist:14:123:decod}
\end{figure*}

\begin{Example}\label{exmp:dist:14:123}
Consider $K=2$ sources with $\T_1 = \{1,4\}$, and $\T_2 = \{1,2,3\}$. The index set of all computable composite messages is
$$\P' = \{\{1\}, \{4\}, \{1,4\}, \{2\}, \{3\}, \{1,2\}, \{1,3\}, \{2,3\}, \{1,2,3\}\}.$$
Similar to Example \ref{exmp:central:1:RR} we set $\K_1 = \{1\}, \K_2 = \{1,2\}, \K_3 = \{3,4\},\K_4 = \{1,4\}$ and obtain the following set of constraints
\begin{align}\label{eq:dist:RR:example2}
R_1&\le S_{1}+S_{1,4},\\R_2 &\le S_2+S_{1,2,3}\\R_1+R_2&\le S_{1}+S_{1,4}+S_2+S_{1,2,3}\\
R_3&\le S_3+S_{1,2,3},\\ R_4 &\le S_{4}+S_{1,4}\\R_3+R_4&\le S_{3}+S_4+S_{1,4}+S_{1,2,3}\\R_1+R_4 &\le S_{1}+S_{1,4}+S_4
\end{align}
To obtain the constraints on rates $S_\J$, we proceed as follows. First, we fix $P(x_1,x_2) = 1/4$ for all $x_1, x_2$. Since receiver 1 knows $S_4$, the MAC constraints are
$S_2+S_3+S_{1,2,3} \le 1$, $S_{1,4} \le 1$ and $S_1+S_{1,4}+S_2+S_3+S_{1,2,3} \le 1.5$. Note that since $W_1$ is common between both sources, the constraint $S_1\le 1$ does not apply. Overall, we obtain the following effective constraints
\begin{align}\label{eq:dist:RR:example111}
S_1+S_2+S_3+S_{1,2,3} +S_{1,4}&\le 1.5\\
S_2+S_3+S_{1,2,3}&\le 1\\
S_{3}+S_{1,2,3}+S_{4}+S_{1,4}&\le 1.5\\
S_4 + S_{1,4} &\le 1\\
S_{1}+S_{1,4}+S_4 &\le 1.5
\end{align}


After Fourier-Motzkin elimination we obtain
\begin{align}
R_2 &\le 1,\quad \quad R_3\le 1,\quad \quad R_4\le 1, \\
R_1+R_2 &\le 1.5,\quad
R_1+R_3 \le 1.5,\\
R_3+R_4 &\le 1.5,\quad R_1+R_4 \le 1.5
\end{align}
Rates $R_1 = R_3 = 0.5$ and $R_2=R_4 = 1$ and hence sum rate $R_1+R_2+R_3+R_4 = 3$ is achievable by setting $X_1 = W_{1,4} = M_1\oplus M_4$ and $X_2 = W_{1,2,3} = M_1\oplus M_2 \oplus M_3$. 

As shown in Fig. \ref{fig:dist:14:123:decod}, it can be verified that when $Y = 0$ (and hence $X_1 = X_2 = 0$), then receivers 1, 2, 3, and 4 decode $M_1 = M_4$, $M_2 = M_3\oplus M_4$, $M_3 = M_1\oplus M_2$ and $M_4 = M_2 \oplus M_3$, respectively. When $Y = 2$ (and hence $X_1 = X_2 = 1$), then receivers 1, 2, 3, and 4 decode $M_1 = M_4\oplus 1$, $M_2 = M_3\oplus M_4$, $M_3 = M_1\oplus M_2 \oplus 1$ and $M_4 = M_2 \oplus M_3$, respectively. When $Y = 1$, then receivers 2 and 4 decode $M_2 = M_3\oplus M_4\oplus 1$ and $M_4 = M_2+M_3 \oplus 1$, respectively. Effectively, receivers 2 and 4, can suppress unknown message $M_1$ be decoding $X_1 \oplus X_2$ from $Y$ - which may be thought of as some form of non-unique decoding. However, when $Y=1$ then $M_1 = M_3 = E$, where $E$ stands for erasure.

\end{Example}

The above examples highlight the impact of distribution of messages across storage nodes on the achievable index coding rates. It is important that messages are distributed such that ``key'' composite messages can be generated and increased sum rate through MAC can be fully utilized. While compared to Example \ref{exmp:central:1:RR},  the sum rate has increased in Examples \ref{exmp:dist:14:23} and \ref{exmp:dist:14:123} due to MAC and repetition (and hence more storage cost in Example \ref{exmp:dist:14:123}), we note that unlike centralized index coding Example $\ref{exmp:central:1:RR}$ composite message $S_{1,2,3,4}$ cannot be generated in the network. In the following example, we show how we can address this limitation and symmetrically increase all rates without actually increasing storage space.

\begin{figure*}[t]
\centering
\includegraphics[width=\linewidth]{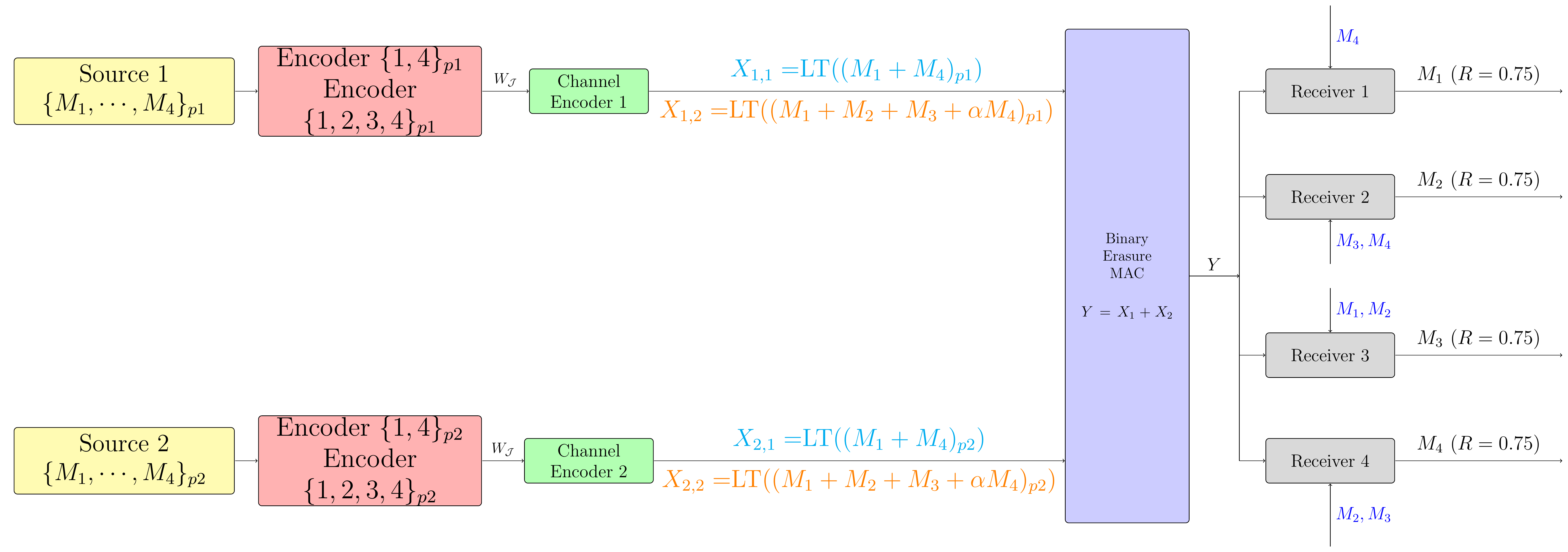}
\caption{The overall schematic of decoding of striped messages for Example \ref{exmp:dist:striping}. By time sharing between decoding $X_1$ first or $X_2$ first in $Y$, a symmetric rate of $R=0.75$ is achievable for all receivers.}
\label{fig:dist:striping:2}
\end{figure*}

\begin{Example}\label{exmp:dist:striping}
Consider $K=2$ sources with ``striped'' message sets $\T_1 = \{1_{p1},2_{p1},3_{p1},4_{p1}\}$, and $\T_2 = \{1_{p2},2_{p2},3_{p2},4_{p2}\}$, where $p1$ and $p2$ stand for parts (halves) 1 and 2 of each message, respectively, and $M_{j_{pk}} \in [1:2^{nR_{j_{pk}}}]$, $j \in [1:N]$, $k = 1,2$ with corresponding rates $R_{j_{pk}}$ and $R_j = R_{j_{p1}}+R_{j_{p2}}$. In this way, required partial composite messages $S_{1_{pk},4_{pk}}$ and $S_{1_{pk},2_{pk}, 3_{pk}, 4_{pk}}$ can be generated at the corresponding source $k$.  This is shown in Fig. \ref{fig:dist:striping:2}. Therefore, following similar steps as in Example \ref{exmp:central:1:RR}, it can be verified that
\begin{align}\label{eq:dist:RR:example3:1}
R_{1_{pk}}+R_{2_{pk}} &\le 1,\quad R_{1_{pk}}+R_{3_{pk}} \le 1\\\label{eq:dist:RR:example3:2}
R_{1_{pk}}+R_{4_{pk}} &\le 1, \quad R_{3_{pk}}+R_{4_{pk}} \le 1
\end{align}
for $k=1,2$. Moreover, due to MAC constraints, the sum of any two inequalities one with $k=1$ and the other with $k=2$ is upper bounded by $I(X_1,X_2;Y) = 1.5$. Crucially, we can verify that we can symmetrically achieve
\begin{align}\label{eq:central:RR:example:striping}
R_1+R_2 &\le 1.5,\quad R_1+R_3 \le 1.5\\
R_1+R_4 &\le 1.5, \quad R_3+R_4 \le 1.5
\end{align}
Compared to Example \ref{exmp:dist:14:123}, the limits of $R_2, R_3, R_4 \le 1$ no longer exist.

\end{Example}

\begin{Remark}\label{rem:1}
Generalizing the above example, we make an important observation. For $K$ sources and $N$ messages, ``striping'' or dividing all messages to $K$ sub-messages and storing each striped set at one source will ensure that all striped composite messages are computable  in the network.  Assuming a symmetric MAC (such as the considered binary erasure MAC), if sum rate $\sum_{j \in \J} R_j < I_\J$ is achievable in the centralized index coding problem with capacity link $I(X_1;Y|X_2, \cdots, X_K)$, then $\sum_{k=1}^{K}\sum_{j \in \J} R_{j_{pk}} < I_\J \times I(X_{[1:K]};Y)$ is achievable in the distributed case. This separable or multiplicative rate region expansion of striped storage means that there will be no loss in index coding rate region due to distributed storage. Nevertheless, there may be diminishing returns in increasing $K$ to the saturation of MAC capacity. For example, by increasing $K=3$ and striping messages into three parts, the sum rate increases to $R_1+R_2+R_3+R_4 = 2 \times I(X_1,X_2,X_3;Y) \approx 2\times 1.81$; around 20\% increase compared to $K=2$.
\end{Remark}

Finally, we consider the effect of coded storage on the achievable rates of distributed index coding. 

\begin{figure*}[t]
\centering
\includegraphics[width=\linewidth]{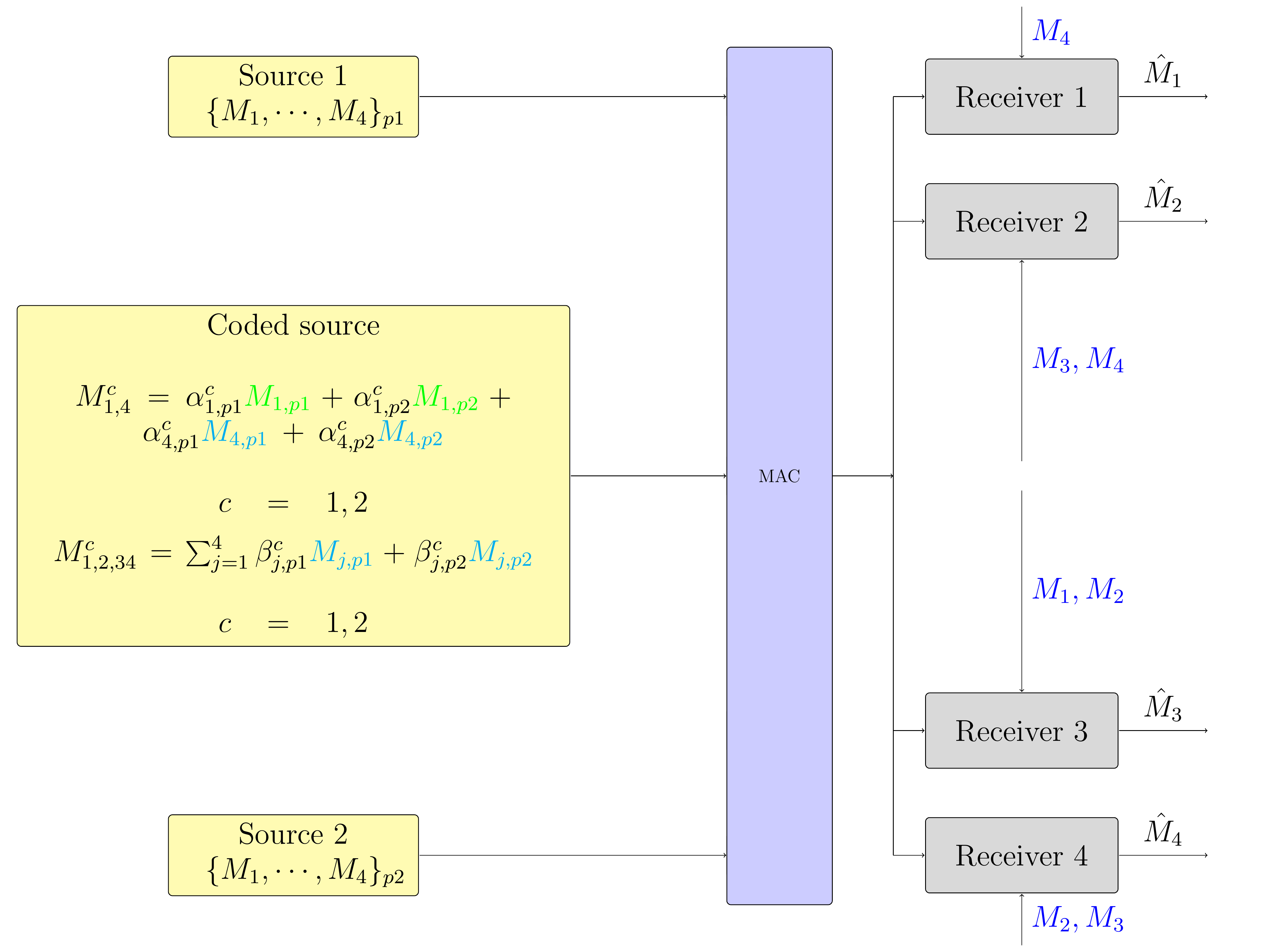}
\caption{The system of coded messages for Example \ref{exmp:dist:MDS:3:2}.}
\label{fig:dist:mds:2}
\end{figure*}

\begin{Example}\label{exmp:dist:MDS:3:2}
Building on Example \ref{exmp:dist:striping}, consider $K=3$ sources with $\T_1 = \{1_{p1},2_{p1},3_{p1},4_{p1}\}$, and $\T_2 = \{1_{p2},2_{p2},3_{p2},4_{p2}\}$ and a ``coded'' source which has access to two ``precoded'' composite messages of the form 
$$W^c_{1,4} = \alpha^c_{1_{p1}} M_{1_{p1}}+\alpha^c_{1_{p2}} M_{1_{p2}}+\alpha^c_{4_{p1}} M_{4_{p1}}+\alpha^c_{4_{p2}} M_{4_{p2}}$$ for $c=1,2$ and two precoded messages of the form 
$$W^c_{1,2,3,4} = \sum_{j=1}^4 \beta^c_{j_{p1}} M_{j_{p1}}+\beta^c_{j_{p2}} M_{j_{p2}}$$ for $c=1,2$. In other words, coded storage has access to pre-coded forms of $S^c_{1,4}$ and $S^c_{1,2,3,4}$, where superscript $c$ stands for precoded composite message. The messages and coefficients are chosen from a finite field $\mathbb F_q$ such that the three sources constitute a (3,2) MDS code which can tolerate any one source failure. A schematic of source messages is shown in Fig. \ref{fig:dist:mds:2}.

Denoting by $S^{p_k}_{1,4}$, $S^{p_k}_{1,2,3,4}$ partial composite messages that can be computed at sources $k=1$ and $k=2$, the essential dual index coding constraints are
\begin{align*}
R_1&\le S^{p_1}_{1,4}+S^{p_2}_{1,4}+S^{c}_{1,4},\\R_2 &\le S^{p_1}_{1,2,3,4}+S^{p_2}_{1,2,3,4}+S^{c}_{1,2,3,4}\\\nonumber R_1+R_2 &\le S^{p_1}_{1,4}+S^{p_2}_{1,4}+S^{c}_{1,4}+S^{p_1}_{1,2,3,4}+S^{p_2}_{1,2,3,4}\\\nonumber&\quad+S^{c}_{1,2,3,4},\\
R_3 &\le S^{p_1}_{1,2,3,4}+S^{p_2}_{1,2,3,4}+S^{c}_{1,2,3,4}\\\nonumber R_3+R_4 &\le S^{p_1}_{1,4}+S^{p_2}_{1,4}+S^{c}_{1,4}+S^{p_1}_{1,2,3,4}+S^{p_2}_{1,2,3,4}\\\nonumber&\quad+S^{c}_{1,2,3,4},\\
R_1+R_4 &\le S^{p_1}_{1,4}+S^{p_2}_{1,4}+S^{c}_{1,4}+S^{p_1}_{1,2,3,4}+S^{p_2}_{1,2,3,4}\\\nonumber&\quad+S^{c}_{1,2,3,4},
\end{align*}
And the set of MAC inequalities are
\begin{align*}
S^{p_k}_{1,4}+S^{p_k}_{1,2,3,4}&\le 1\\
S^{c}_{1,4}+S^{c}_{1,2,3,4}&\le 1\\
\sum_{k=1}^2S^{p_k}_{1,4}+S^{p_k}_{1,2,3,4}&\le 1.5\\
S^{p_k}_{1,4}+S^{p_k}_{1,2,3,4}+S^{c}_{1,4}+S^{c}_{1,2,3,4}&\le 1.5\\
S^{p_1}_{1,4}+S^{p_2}_{1,4}+S^{c}_{1,4}+S^{p_1}_{1,2,3,4}+S^{p_2}_{1,2,3,4}+S^{c}_{1,2,3,4} &\le 1.81
\end{align*}
where after Fourier-Motzkin elimination results in
\begin{align}\label{eq:exmp:dist:MDS:3:2}
R_1+R_2 &\le 1.81,\quad R_1+R_3 \le 1.81\\
R_1+R_4 &\le 1.81, \quad R_3+R_4 \le 1.81
\end{align}
Therefore, this example shows that the sum rate of $3.62$ is possible as it is with normal striping of messages into $K=3$ under Remark \ref{rem:1}, but with the added advantage of resilience to one source node failure. The symmetric rates $R_1=R_2=R_3 = R_4 = 0.31/2+0.5/2+1/2 = 0.9050$ can be achieved if striped sources $k=1,2$ send striped composite messages
$$W^k_{1,4} = \gamma^k_{1_{pk}} M_{1_{pk}}+\gamma^k_{4_{pk}} M_{4_{pk}}$$ 
and
$$W^k_{1,2,3,4}= \sum_{j=1}^4 \delta^k_{j_{pk}} M_{j_{pk}}$$
alternately and the coded source $k=3$ sends it 4 pre-coded messages $W^c_{1,4}$ and $W^c_{1,2,3,4}$ for $c=1$ or $c = 2$ alternately. 
\end{Example}

\newpage

\end{document}